\documentclass[american,english,british,aps,manuscript,prl,final,showpacs,lengthcheck]{revtex4-1}
\usepackage[T1]{fontenc}
\usepackage[latin9]{inputenc}
\usepackage{color}
\usepackage{amsmath}
\usepackage{amssymb}
\usepackage{graphicx}
\usepackage{esint}

\makeatletter
 
 \@ifundefined{textcolor}{}
 {%
   \definecolor{BLACK}{gray}{0}
   \definecolor{WHITE}{gray}{1}
   \definecolor{RED}{rgb}{1,0,0}
   \definecolor{GREEN}{rgb}{0,1,0}
   \definecolor{BLUE}{rgb}{0,0,1}
   \definecolor{CYAN}{cmyk}{1,0,0,0}
   \definecolor{MAGENTA}{cmyk}{0,1,0,0}
   \definecolor{YELLOW}{cmyk}{0,0,1,0}
 }

\usepackage{babel}
\usepackage{pdfpages}

\makeatother

\usepackage{babel}
\begin{document}

\title{Dynamical control of state transfer through noisy quantum channels:
optimal tradeoff of speed and fidelity}

\author{Analia Zwick, Gonzalo A. Alvarez, Guy Bensky and Gershon Kurizki}

\affiliation{Weizmann Institute of Science, Rehovot 76100, Israel}
\begin{abstract}
We propose a method of optimally controlling state transfer through
a noisy quantum channel (spin-chain). This process is treated as qubit
state-transfer through a fermionic bath. We show that dynamical modulation
of the boundary-qubits levels suffices to ensure fast and high-fidelity
state transfer. This is achievable by dynamically optimizing the transmission
spectrum of the channel. The resulting optimal control is robust against
both static and fluctuating noise. 
\end{abstract}
\maketitle
One dimensional (1D) chains of spin-$\frac{1}{2}$ systems with nearest-neighbor
couplings, nicknamed spin chains, constitute a paradigmatic quantum
many-body system of the Ising type \cite{ising_beitrag_1925,bethe_1931},
whose treatment is nontrivial yet manageable. As such, spin chains
are well suited for studying the transition from quantum to classical
transport and from mobility to localization of excitations as a function
of disorder and temperature \cite{kramer_localization:_1993}. In
the context of quantum information (QI), spin chains are envisioned
to form reliable quantum channels for QI transmission between nodes
(or blocks) of quantum communication or coupling schemes \cite{bose_quantum_2003}.
Contenders for the realization of high-fidelity QI transmission are
spin chains comprised of superconducting qubits \cite{lyakhov_quantum_2005,majer_coupling_2007},
cold atoms \cite{duan_controlling_2003,hartmann_effective_2007,fukuhara_quantum_2013,simon_quantum_2011},
nuclear spins in liquid- or solid-state NMR \cite{madi_time-resolved_1997,zhang_simulation_2005,cappellaro_dynamics_2007,zhang_iterative_2007,rufeil-fiori_effective_2009,alvarez_perfect_2010,ajoy_algorithmic_2012}
quantum dots \cite{nikolopoulos_electron_2004,*petrosyan_coherent_2006},
ion traps \cite{lanyon_universal_2011,blatt_quantum_2012} and nitrogen-vacancy
(NV) centers in diamond \cite{cappellaro_coherence_2009,neumann_quantum_2010,yao_scalable_2012,ping_practicality_2013}.

The distribution of coupling strengths between the spins that form
the quantum channel, determines the state transfer-fidelities \cite{bose_quantum_2003,christandl_perfect_2005,kay_perfect_2006,*kay_perfect_2010,karbach_spin_2005}.
Perfect state-transfer (PST) channels can be obtained by precisely
engineering each of those couplings. Such engineering is however highly
challenging at present \cite{cappellaro_dynamics_2007,zwick_robustness_2011}.
A much simpler control may involve \textit{only} the boundary (source
and target) qubits that are connected via the channel. Recently, it
has been shown that if the boundary qubits are weakly-coupled to a
uniform (homogeneous) channel (\textit{i.e.}, one with identical couplings),
quantum states can be transmitted with arbitrarily high fidelity at
the expense of increasing the transfer time \cite{wojcik_unmodulated_2005,*wojcik_multiuser_2007,zwick_quantum_2011,yao_robust_2011,zwick_spin_2012}.
Yet such slowdown of the transfer may be detrimental because of omnipresent
decoherence. To overcome this problem, we here propose a hitherto
unexplored approach for optimizing the tradeoff between fidelity and
speed of state-transfer in quantum channels. This approach employs
temporal modulation of the couplings between the boundary qubits and
the rest of the channel, which is treated as dynamical control of
a quantum system coupled to a fermionic bath. The goal of the modulation
is to realize an optimal spectral filter \cite{clausen_bath-optimized_2010,*clausen_task-optimized_2012,escher_optimized_2011,*bensky_optimizing_2012,*petrosyan_reversible_2009, gordon_universal_2007,*gordon_optimal_2008,kofman_universal_2001,*kofman_unified_2004}
that blocks transfer via the eigenmodes of the channel that are responsible
for leakage of the QI \cite{wu_master_2009}. We show that under optimal
modulation, the fidelity and the speed of transfer can be improved
by several orders of magnitude, and the fastest transfer is achievable
for a given fidelity .

\textit{Quantum channel: Hamiltonian and boundary control.---} In
keeping with previous studies \cite{bose_quantum_2003,lyakhov_quantum_2005,majer_coupling_2007,duan_controlling_2003,hartmann_effective_2007,fukuhara_quantum_2013,madi_time-resolved_1997,zhang_simulation_2005,cappellaro_dynamics_2007,zhang_iterative_2007,simon_quantum_2011,rufeil-fiori_effective_2009,alvarez_perfect_2010,cappellaro_coherence_2009,neumann_quantum_2010,yao_scalable_2012,christandl_perfect_2005,kay_perfect_2006,*kay_perfect_2010,karbach_spin_2005,zwick_robustness_2011,wojcik_unmodulated_2005,zwick_quantum_2011,yao_robust_2011,zwick_spin_2012,ajoy_algorithmic_2012,nikolopoulos_electron_2004,petrosyan_coherent_2006,ping_practicality_2013,blatt_quantum_2012,lanyon_universal_2011},
we consider a spin-$\frac{1}{2}$ chain with XX interactions between
nearest neighbors. The Hamiltonian is given by 
\begin{equation}
\begin{array}{c}
H=H_{0}+H_{bc}(t),\\
H_{0}=\frac{J_{i}}{2}\sum_{i=1}^{N-1}\left(\sigma_{i}^{x}\sigma_{i+1}^{x}+\sigma_{i}^{y}\sigma_{i+1}^{y}\right)\\
H_{bc}(t)=\frac{J_{i}}{2}\alpha(t)\sum_{i=0,N}\left(\sigma_{i}^{x}\sigma_{i+1}^{x}+\sigma_{i}^{y}\sigma_{i+1}^{y}\right)
\end{array},\label{eq:hamiltonian-1}
\end{equation}
where $H_{0}$ and $H_{bc}$ stand for the chain and boundary-coupling
Hamiltonians, respectively, $\sigma_{i}^{\mu}$ are the Pauli matrices,
$N$ is the chain length, and $J_{i}>0$ is the exchange interaction
coupling. \foreignlanguage{english}{The magnetization-conserving $H_{0}$
can be transformed into a non-interacting fermionic Hamiltonian \cite{lieb_two_1961},
that has the diagonal, particle-conserving form $H_{0}=\sum_{k=1}^{N}\omega_{k}b_{k}^{\dagger}b_{k}$,}\foreignlanguage{american}{
where $b_{k}^{\dagger}$}\foreignlanguage{english}{ populates a fermionic
single-particle, eigenstate of energy}\foreignlanguage{american}{
$\omega_{k}$.}

Under the assumption of mirror symmetry of the couplings $J_{i}\!=\!J_{N-i}$
for odd $N$, there is a single non-degenerate, zero-energy fermionic
mode in the quantum channel \cite{kay_perfect_2006,*kay_perfect_2010,karbach_spin_2005,zwick_robustness_2011}, corresponding to $k\!=\!z\!=\!\frac{N+1}{2}$.
The two boundary qubits are resonantly coupled to this mode \cite{wojcik_multiuser_2007,yao_robust_2011,ping_practicality_2013} with an
effective, temporally-modulated coupling strength $\tilde{J}_{z}\alpha(t)$.
This resonant fermionic tunneling is described by the effective Hamiltonian
\begin{equation}
H_{S}(t)=\tilde{J}_{z}\alpha(t)(c_{0}^{\dagger}b_{z}+c_{N+1}^{\dagger}b_{z}+\mathrm{h.c.}).
\end{equation}

The main idea of our treatment is to consider these three fermionic
modes as a system $S$ that interacts with a bath $B$, and thus rewrite
the total Hamiltonian as $H=H_{S}(t)+H_{B}+H{}_{SB}(t)$\foreignlanguage{english}{,
where $H_{B}=\sum_{k=1}^{N}\omega_{k}b_{k}^{\dagger}b_{k}$ with $k\neq z,\, k=1...N$.
Upon defining the collective-mode operators $\tilde{b_{k}}_{odd(even)}=\sum_{k_{odd(even)}=1}^{N}\tilde{J}_{k}b_{k},$
the system-bath interaction assumes the form 
\begin{equation}
H_{SB}(t)\!=\!\alpha(t)[(c_{0}^{\dagger}+c_{N+1}^{\dagger})\tilde{b}_{k_{odd}}+(c_{0}^{\dagger}-c_{N+1}^{\dagger})\tilde{b}_{k_{even}}]+\mathrm{h.c.}\label{eq:Hsb}
\end{equation}
This form is amenable to optimal dynamical control of the multipartite
system \cite{clausen_bath-optimized_2010,*clausen_task-optimized_2012,gordon_scalability_2011,*gordon_dynamical_2009,*kurizki_universal_2013}
that generalizes single-qubit dynamical control by modulation of the
qubit levels }\cite{escher_optimized_2011,*bensky_optimizing_2012,*petrosyan_reversible_2009,gordon_universal_2007,*gordon_optimal_2008,kofman_universal_2001,*kofman_unified_2004}\foreignlanguage{english}{.}

\begin{figure}
\centering{}\includegraphics[width=0.99\columnwidth]{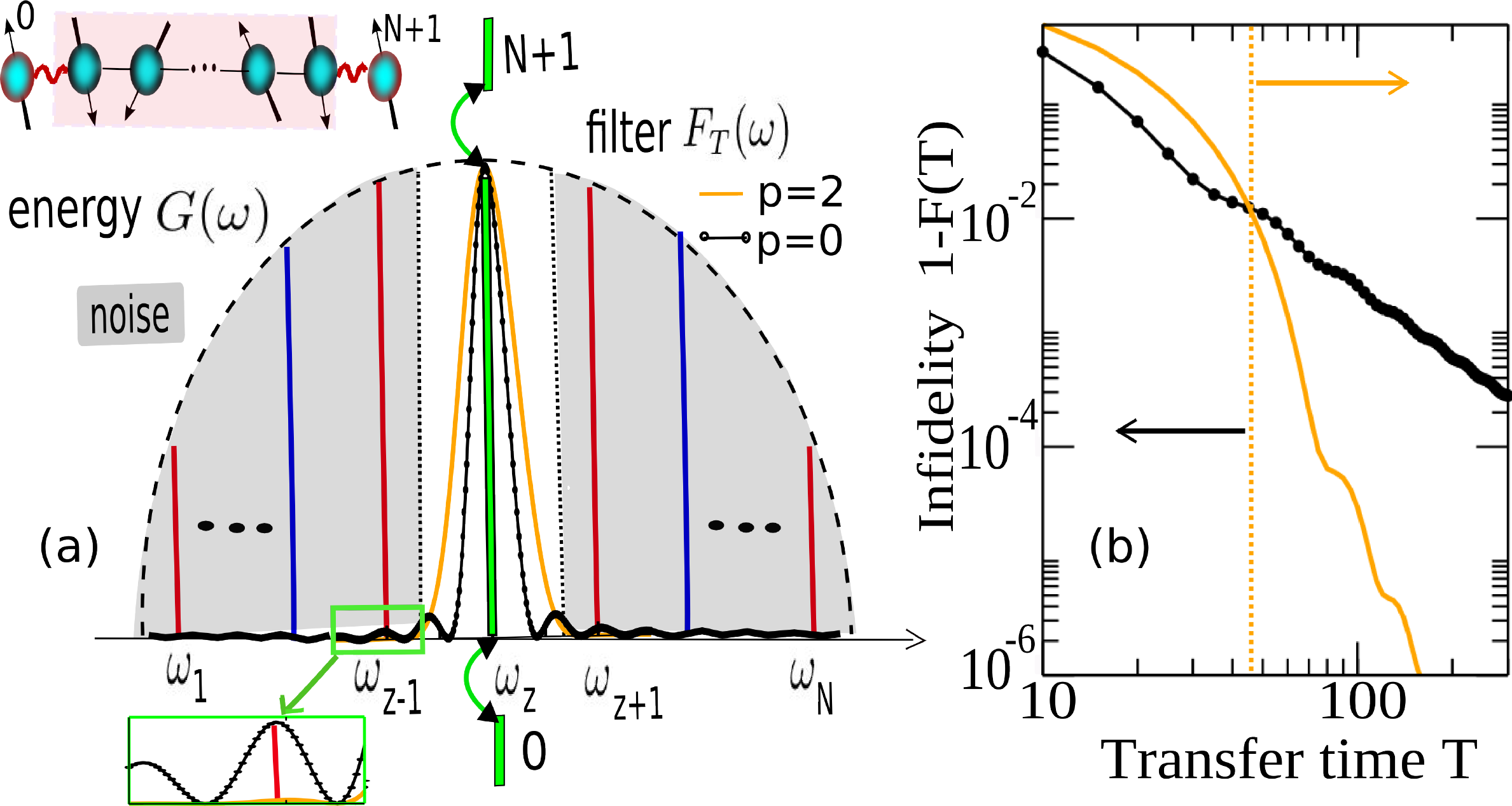}\caption{\label{fig:chain-FilterFunction}(Color online) Top inset: Spin-channel
for state transfer with boundary-controlled couplings. Boundary-controlled
spin chain mapped to a non-interacting spinless fermions system. The
two boundary spins 0 and $N+1$ are resonantly coupled to the chain
by the fermionic-mode $z$ with a coupling strength $\tilde{J}_{z}\alpha(t).$
(a) Spectrum of the effective fermionic system (rectangular bars)
which interacts with the bath-modes $k$ (red-even $k$ and blue-odd
$k$ vertical lines) with strengths $\tilde{J}_{k}\alpha(t)$. Dashed
contour: noise spectrum described by the Wigner-semicircle (maximal-disorder)
lineshape with a central gap around $\omega_{z}$. In the central
gap, the optimal spectral-filters $F_{T}(\omega)$ generated by dynamical
boundary-control with $\alpha_{p}(t)$ $(p=0\mbox{ (black dotted), }p=2\mbox{ (orange thin)})$
are shown. Bottom inset: a zoom of the tails of the filter spectrum
that protect the state transfer against a general noisy bath with
a central gap. (b) Infidelity as a function of transfer time $T$
under optimal control (filter) with $p=0$ (black dotted) and $p=2$
(orange thin).}
\end{figure}

To this end, we rewrite Eq. (\ref{eq:Hsb}) in the interaction picture
as $H_{SB}^{I}(t)=\sum_{j}S_{j}(t)\otimes B_{j}^{\dagger}(t)$\foreignlanguage{american}{;}\foreignlanguage{english}{
and decompose $H_{SB}^{I}(t)$ into symmetric and antisymmetric system
operators that are coupled to odd- and even-bath modes (see SI).}
Upon representing the system operators $S_{j}(t)$ via a rotation-matrix
$\Omega_{j,i}(t)$ in a chosen basis of operators $\hat{\nu}_{i}$,
so that\foreignlanguage{english}{ $S_{j}(t)=\sum_{i}\Omega_{j,i}(t)\hat{\nu}_{i},$
we can write a time-convolutionless second-order solution for the
system density matrix $\rho_{S}(t)$ in the interaction picture \cite{clausen_bath-optimized_2010,*clausen_task-optimized_2012}.
This solution will be used to calculate and optimize the transfer
fidelity in what follows.}

\selectlanguage{english}%
Let us consider a generic qubit-state\foreignlanguage{british}{ }$\vert\psi_{0}\rangle=\alpha\vert0_{0}\rangle+\beta\vert1_{0}\rangle$\foreignlanguage{british}{
as the source qubit $0$, }and $\vert\psi\rangle_{S}\otimes\vert0\rangle_{B}$
with $\vert\psi\rangle_{S}=\vert\psi_{0}\rangle\otimes\vert0_{z}0_{N+1}\rangle{}_{S}$
as the initial state of $S+B$. \foreignlanguage{british}{We shall
be interested in the transfer fidelity }of $\vert\psi_{0}\rangle$
to the target qubit $N+1$, averaged over all input states on the
Bloch sphere: it is given by \cite{bose_quantum_2003} $F(T)=\frac{f_{0,N+1}^{2}(T)}{6}+\frac{f_{0,N+1}(T)}{3}+\frac{1}{2}$,
where $f_{0,N+1}(T)$ is simply the transfer fidelity of $\vert\psi_{0}\rangle=\vert1_{0}\rangle$.
This transfer fidelity \foreignlanguage{american}{is expressed in
the interaction picture as $f_{0,N+1}(T)=\left|_{S}\left\langle \psi\right|\rho_{S}(T)\left|\psi\right\rangle _{S}\right|$}
\foreignlanguage{american}{where}\foreignlanguage{british}{ $T$ is
}the transfer time\foreignlanguage{british}{.} The transfer fidelity
remains the same for any initial state \foreignlanguage{british}{of
the bath channel $B$ withing the weak coupling regime \cite{danieli_quantum_2005,yao_robust_2011,ping_practicality_2013}.}

\selectlanguage{british}%
\textit{Optimization method.---} To ensure the best possible state-transfer
fidelity, we use modulation as a tool to minimize\foreignlanguage{american}{
the infidelity $\zeta(T)=1-f_{0,N+1}(T)$ by rendering} the overlap
between the bath and system spectra\foreignlanguage{english}{ as small
as possible (see SI )\cite{clausen_bath-optimized_2010,*clausen_task-optimized_2012}.}
The infidelity may be written as the convoluted overlap\foreignlanguage{english}{
\begin{equation}
\zeta(T)\!=\!\Re\int_{0}^{T}\!\! dt\!\int_{0}^{t}\! dt'\!\!\!\!\sum_{q=even,odd}\!\!\!\!\Omega_{q}(t)\Omega_{q}(t')\Phi_{q}(t-t')\label{eq:eta_timedomain}
\end{equation}
}where $\Phi_{odd(even)}(\tau)=\sum_{k{}_{odd(even)}}|\tilde{J}_{k}|^{2}e^{-i\omega_{k}\tau}$
are the \foreignlanguage{english}{bath-correlation functions}, while
$\Omega_{odd}(\tau)=\dot{\phi}(\tau)cos(\sqrt{2}\phi(\tau))/\tilde{J}_{Z}^{2}$
and $\Omega_{even}(\tau)=\dot{\phi}(\tau)/\tilde{J}_{Z}^{2}$ are
the dynamical control functions, \foreignlanguage{english}{expressed
in terms of the }the phase accumulated by the qubit under modulation
control $\phi(T)=\tilde{J_{z}}\int_{0}^{T}\alpha(t')dt'$. In the
energy domain, Eq. (\ref{eq:eta_timedomain})\foreignlanguage{english}{
has the form $\zeta(T)=\underset{q=even,odd}{\overset{}{\sum}}\int G^{q}(\omega)F_{T}^{q}(\omega)d\omega$,}
where the Fourier transforms $G^{q}(\omega)=\mathcal{FT}(\Phi_{q}(\tau))$
and $F_{T}^{q}(\omega)=\mathcal{FT}(\frac{|\Omega_{q}(t)|^{2}}{2\pi})$
are the bath-spectrum and the filter-energy functions, respectively,
for even or odd $q$. To determine the optimal modulation control,
we minimize the overlap integrals of $G^{q}(\omega)$ and $F_{T}^{q}(\omega)$
for a given $T$ by the variational Euler-Lagrange method.

\selectlanguage{english}%
We require the channel to be symmetric with respect to the source
and target qubits and the number of eigenvalues to be odd. This requirements
allows for a central eigenvalue that is invariant under noise on the
couplings, provided a gap exists between this eigenvalue and the adjacent
ones, \textit{i.e.} they are not strongly blurred (mixed) by noise,
so as not to make them overlap. The optimized modulations derived
here are applicable to any system of this kind. As an example, consider
a uniform (homogeneous) spin-chain channel, i.e. $J_{i}\equiv J$
and energies \foreignlanguage{american}{$\omega_{k}=2Jcos(\frac{k\pi}{N+1})$.}
In this case, under complete randomization of $\omega_{k}$, the lineshape
is the Wigner semicircle (see \ref{fig:chain-FilterFunction}a and
SI). In the weak-coupling regime $\alpha\ll1$, the interaction $H_{bc}$
is treated perturbatively, so that $\tilde{J}_{z}=\sqrt{\frac{2}{N+1}}J$
and $\tilde{J}_{k}=\tilde{J}_{z}sin(\frac{k\pi}{N+1})$ are always
much smaller than the nearest eigenvalue gap $\vert\omega_{z}-\omega_{z\pm1}\vert\sim\frac{2J}{N}$
\cite{wojcik_unmodulated_2005,wojcik_multiuser_2007,yao_robust_2011}.
This may not happen in the strong coupling regime $\alpha\sim1$,
which requires special consideration (see below).

\selectlanguage{british}%
\textit{Optimal filter design.---} In order to obtain universal solutions
for channels with static or fluctuating spin-spin coupling noise,
we assume that the discreteness of the quantum channel spectrum is
smoothed out by the noise (see SI). On the other hand, the noise-induced
broadening is assumed to be lower than the gap around the eigenvalue
$\omega_{z}=0$ that remain invariant against this noise, whereas
higher eigenvalues are affected by it \cite{zwick_robustness_2011}.
A filter that is efficient and robust against noise in a system with
a central gap must be a narrow bandpass around $\omega_{z}$. To this
end, we look for maximized $F_{T}(\tau)=\int F_{T}(\omega)e^{-i\omega\tau}d\omega$
for every $\tau$, in order to ensure that only the lowest frequency
components are present in the filter-energy function under the constraint
of accumulated phase $\phi(T)$ and energy $E(T)=\tilde{J}_{z}^{2}\int_{0}^{T}|\alpha(t)|^{2}dt\geq\frac{\phi(T)^{2}}{T}$
\foreignlanguage{english}{\cite{escher_optimized_2011,*bensky_optimizing_2012,*petrosyan_reversible_2009}}.

\selectlanguage{english}%
The optimal solutions are found to be (see SI)
\begin{equation}
\alpha_{p}(t)=\alpha_{M}sin^{p}\left(\frac{\pi t}{T_{p}}\right)\label{eq:Opt Mod alphap}
\end{equation}
with $p=0,1,2$\textcolor{black}{,} $T_{p}=c_{p}\frac{\phi(T)}{J_{z}}$
and $c_{p}=\frac{\sqrt{\pi}\Gamma(\frac{1+p}{2})}{\Gamma(\frac{1+p}{2})}$
($c_{0}=1,\, c_{1}=\frac{\pi}{2},\, c_{2}=2$). Here $p=0$ means
static control, while $p=1,2$ stands for dynamical control. For $p=0$,
$\alpha_{0}(t)$ is constant and satisfies the minimal-energy condition,
$E_{min}(T_{0})=\frac{\pi^{2}}{2T_{0}}$.

Although the corresponding filter function is \foreignlanguage{british}{a
narrow bandpass} around $0$, it still has many wiggles (Fig. \ref{fig:chain-FilterFunction}a)
which overlap with bath-energies that hamper the transfer. Therefore,
to improve the fidelity transfer we require a filter that is flatter
and lower throughout the bath-energy domain. By allowing $E\gtrsim E_{min}$,
the filter is made lower outside a region around 0 by the modulation
$\alpha_{1}(t)$, with $E_{1}=\frac{\pi^{2}}{8}E_{min}$, or $\alpha_{2}(t)$,
with $E_{2}=\frac{3}{2}E_{min}$ (Fig. \ref{fig:chain-FilterFunction}a).
This modulation control allows the design of optimal filters $F_{T}^{even}(\omega)$
that are sharp around $0$ and flat (and low) across the bath-energy
range ( $F_{T}^{odd}(\omega)$ filters out the same spectral range).
The inset in Fig. \ref{fig:chain-FilterFunction}a shows that depending
on $T$, different modulations $\alpha_{p}(t)$ are optimal. They
are determined by the overlap between the bath-spectrum, the width
of the central peak and the tail of the filter function. The shorter
is $T$, the lower is $p$ that yields the highest fidelity, because
the central peak that gives the dominant overlap is then the narrowest.
However, as $T$ is increased, larger $p$ will give higher fidelity,
because now the tails give the dominant contribution to the overlap.
As shown in Fig. \ref{fig:chain-FilterFunction}b, the filter for
$p=2$ (similarly for $p=1$) can improve the transfer fidelity by
orders of magnitude in a general noisy gapped-bath.

\begin{figure}
\selectlanguage{british}%
\centering{}\includegraphics[width=0.99\columnwidth]{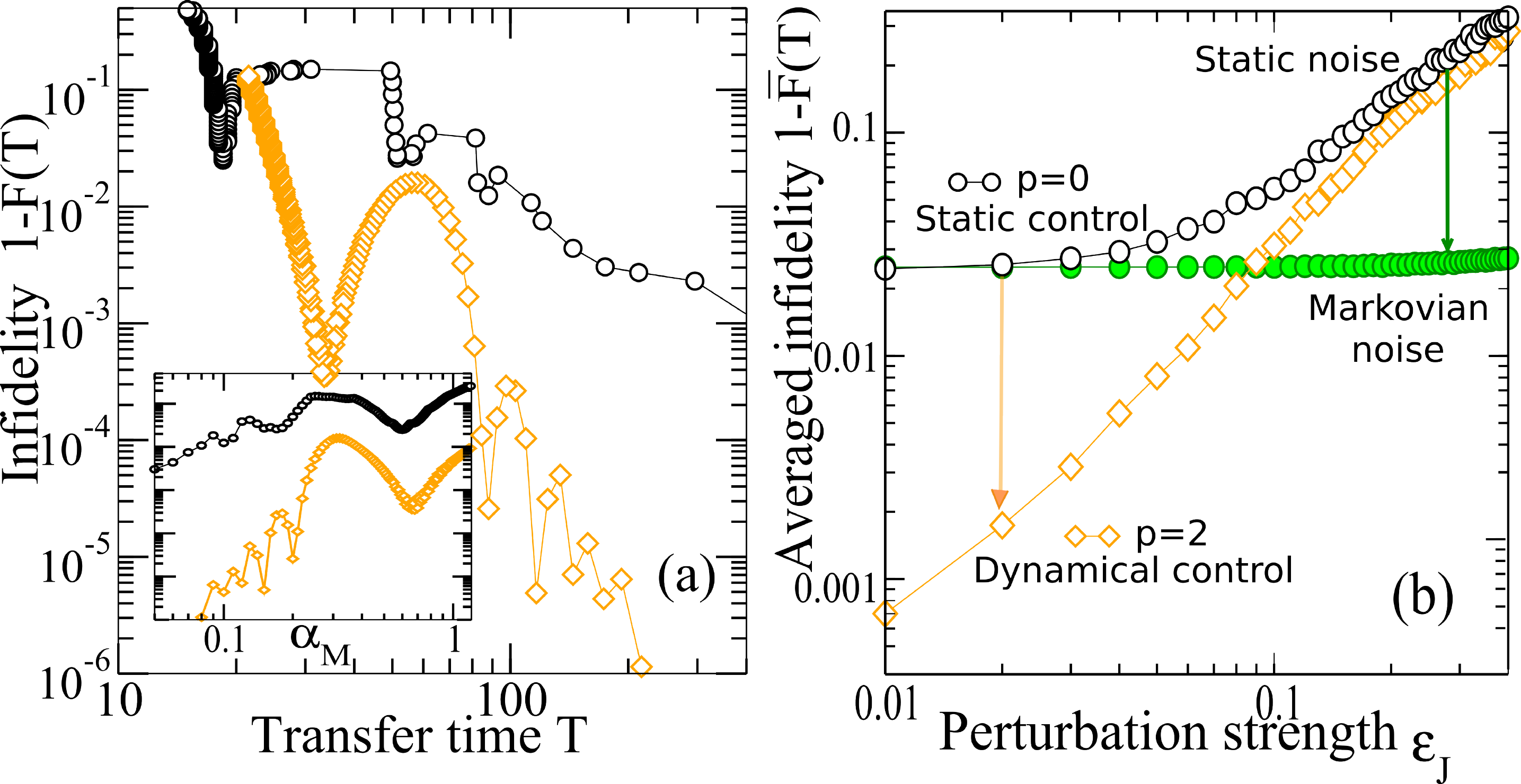}\caption{\label{fig:Fmax-alp}(Color online) Transfer infidelity $1-F(T)$
for a modulated boundary-controlled coupling $\alpha_{p}(t)=\alpha_{M}sin^{p}(\frac{\pi t}{T})$
as a function of (a) the transfer time $T$, (inset) the maximum value
of the boundary coupling $\alpha_{M}$ and (b) the perturbation strength
$\varepsilon_{J}$ of the noisy channel, averaged over \foreignlanguage{american}{$10^{3}$}\foreignlanguage{english}{
noise realizations for $\alpha_{M_{0}}^{opt}=0.6$ and $\alpha_{M_{2}}^{opt}=0.7$.
In static noisy channels, the infidelity obtained under static control
$p=0$ (empty circles) is shown to be strongly reduced under dynamical
$p=2$ control (empty squares). A fluctuating noisy channel is less
damaging; in the Markovian limit where the correlation time of the
noise fluctuations $\tau_{c}\rightarrow0$ ($p=0$, green solid circles),
the infidelity converges to its unperturbed value. $N+2=31$ spins
and $J=1$.}}
\selectlanguage{english}%
\end{figure}

\selectlanguage{british}%
While the \foreignlanguage{english}{approach based on Eq. (\ref{eq:eta_timedomain})}
strictly holds in the weak coupling-regime $(\alpha_{M}\ll1)$\cite{clausen_bath-optimized_2010,*clausen_task-optimized_2012,escher_optimized_2011,*bensky_optimizing_2012,*petrosyan_reversible_2009,kofman_universal_2001,*kofman_unified_2004},
the validity of the optimal modulations can also be extended to strong
couplings $\alpha_{M}$, since they become compatible with the weak-coupling
regime under the filtering process. This is observed, for example,
for a homogeneous channel in Fig. \ref{fig:Fmax-alp}a, where the
state-transfer infidelity is displayed as a function of $\alpha_{M}$
and $T$, for $\alpha_{p}(t)=\alpha_{M}sin^{p}(\frac{\pi t}{T})$
with $p=0,2$. In the weak-coupling regime $(\alpha_{M}\ll1)$ the
infidelity decreases with $\alpha_{M}$ according to a power law,
and the transfer time increases \foreignlanguage{english}{as $T\approx c_{p}\frac{\pi\sqrt{N}}{2\alpha_{M}J}$}.
Under optimal-filtering in the strong coupling regime \cite{zwick_quantum_2011,banchi_long_2011,*banchi_optimal_2010,zwick_spin_2012},
there is a minimum infidelity at $\alpha_{M_{p}}^{opt}$ that depends
of $p$. The corresponding transfer time \foreignlanguage{english}{is
$T\approx c_{p}\frac{N}{2J}$}. Here, the oscillatory behavior of
the infidelity reflects the discrete nature of the spectrum. The filter
tails are sinc-like functions, so that when a zero of the filter matches
a bath-energy eigenvalue, the infidelity exhibits a dip.

\selectlanguage{english}%
While Fig. \foreignlanguage{british}{\ref{fig:Fmax-alp}a shows the
transfer infidelity at time $T$, it is important to note that the
fidelity under optimal modulation, $F(t)$, }yields the widest window
of time where the fidelity remains high compared with the unmodulated
cases. This gives more time for determining the transfered state or
using it for further processing, thus increasing the robustness against
imperfection in the temporal accuracy of the optimal dynamical control.

\selectlanguage{british}%
The advantages of dynamical control ($p=1$ or $2$) of the boundary-couplings
are evident in Fig. \ref{fig:Fmax-alp}a.\foreignlanguage{english}{
The inset panel shows that by fixing $\max(\alpha(t))=\alpha_{M}$,
the dynamical modulation increases the transfer fidelity by orders
of magnitude only at the expense of slowing down the transfer time
at most by a factor of 2. By contrast, without dynamical modulation
($p=0$), the optimal $\alpha_{0}^{opt}$ value yields faster transfer(main
panel), but no significant increase of the fidelity. Namely, the only
option for increasing the fidelity is then to reduce $\alpha$, but
the transfer time then increases as$\backsim\frac{1}{\alpha}$. If
the constraint on $\alpha_{M}$ can be relaxed, i.e. more energy can
be used, the great advantages of dynamical control can be appreciated
in both respects, i.e. fidelity increase and transfer-time reduction
by orders of magnitude. Hence, our main result is that the speed-fidelity
tradeoff can be drastically improved under optimal dynamical control.
In particular, optimized modulations provide the fastest transfer
for a given fidelity.}

\textit{Robustness against different noises.--- }We now consider the
effects of noise affecting the coupling strengths as follows: $J_{i}\rightarrow J_{i}+J_{i}\Delta_{i}(t),\; i=1,...,N$
with $\Delta_{i}(t)$ being a uniformly distributed random variable
in the interval $\left[-\varepsilon_{J},\varepsilon_{J}\right]$ for
a given time $t$. Here, $\varepsilon_{J}>0$ characterizes the strength
of the disorder. When $\Delta_{i}(t)$ is independent of time, we
call it \textit{static noise} (this kind of disorder is considered
in Refs. \cite{de_chiara_perfect_2005,petrosyan_state_2010,zwick_robustness_2011,zwick_spin_2012,ronke_effect_2011}),
otherwise we call it \textit{fluctuating-noise}. The following cases
may be discerned:

\textit{(i) Static-noise}. Static control on the boundary-couplings
can make the channel robust against static noise \cite{zwick_spin_2012}
but here we show that dynamical boundary-control makes the channel
even more robust, because it filters out the bath-energies that damage
the transfer. To illustrate this, we compare in Fig. \ref{fig:Fmax-alp}b
the robustness of modulations $\alpha_{p}(t)$, $p=0$ and 2 in the
strong-coupling regime for $\alpha_{M_{p}}=\alpha_{M_{p}}^{opt}$
where the advantage of $p=2$ compared with the static control case
$p=0$ is evident, at the expense of increasing the transfer time
by only a factor of 2. In the weak-coupling regime we may choose $\alpha_{M_{p}}$
such that the transfer fidelity is similar for $p=0$ and $p=2$,
and both cases are similarly robust under disorder, but the modulated
case $p=2$ is an order of magnitude faster. This speedup will be
important in the presence of other sources of decoherence (see below).
We obtain a bound for the fidelity improvement of the state transfer
that is intrinsic to the channel: because of Anderson localization
\cite{Porter1965,akulin_spectral_1993,*pellegrin_mie_2001,Imry2002},
regardless of how small is $\alpha_{0}$, the fidelity cannot be improved
beyond the bound 
\begin{equation}
1-\bar{F}\sim\frac{1}{5}N\varepsilon_{J}^{2},\:(\varepsilon_{J}\ll1).
\end{equation}
\textit{(ii) Markovian noise.--- }In the limit where the gap-width
goes to zero, \textit{i.e.} for a Markovian noise such that the bath
correlation function vanishes at $t-t'>0$, the optimal modulation
can be approximated by $\alpha(t)\approx\alpha_{M}(a+b\, sin^{p}(\frac{t\pi}{T}))$,
where $p\sim3.5$,\textbf{ $\frac{a}{b}\sim\frac{1}{3}$} and $\alpha_{M}=\max\alpha(t)$.
\foreignlanguage{english}{However, the infidelity} for this optimal
modulation almost coincides with the one obtained without modulation.
Thus, modulation is not helpful in the Markovian limit. Counterintuitively,
arbitrarily high fidelities can be achieved for such a bath by slowing
down the transfer time, \textit{i.e}. by decreasing $\alpha_{M}$.
This comes about because in a Markovian bath, the very fast coupling
fluctuations\foreignlanguage{english}{ generate an effective self-decoupling
of the disorder, thereby suppressing the Anderson localization effects
that hamper the transfer fidelity.}

\textit{(iii) Non-Markovian noise}. We finally consider fluctuating
noise \foreignlanguage{american}{$J_{i}+J_{i}\Delta_{i}(t)$ in a
homogeneous channel with constant boundary-couplings. By reducing
the noise correlation time $\tau_{c}$, we observe a convergence of
the transfer fidelity to its value without noise as $\tau_{c}$ decrease
(Fig. \ref{fig:Fmax-alp}b). }\foreignlanguage{english}{Consequently
the fidelity can be substantially improved by reducing $\alpha_{M}$}\foreignlanguage{american}{.
The effective noise strength scales down as $\tau_{c}^{1/2}$, approaching
the Markovian limit when }\foreignlanguage{english}{$\tau_{c}\rightarrow0$}\foreignlanguage{american}{.
As we saw above, modulation is not helpful in the gapless Markovian
limit. By contrast, in the non-Markovian regime that lies between
the static and Markovian limits and the bath-spectrum is gapped, dynamical
control can strongly reduce the infidelity.}

\textit{Realizations.--} A general procedure applicable to any system
which allows control of the boundary spins, consists in modulatating
the boundary couplings by creating an effective Hamiltonian via Trotter-Susuki
decompositions \cite{alvarez_perfect_2010,alvarez_nmr_2010,lanyon_universal_2011,blatt_quantum_2012}.
The corresponding modulation of the boundary-spins energy is only
required to set them on-and off-resonance intermittently at suitable
times \cite{alvarez_perfect_2010} or modulating the boundary couplings
by sequences of $\pi$-pulses on the boundary spins (see SI). In the
weak-coupling regime, efficient transfer through noisy spin chains
is realizable by periodically modulating the level distance of the
boundary qubits by an off-resonant field, whose effect in this regime
is the same as periodically modulating the qubits coupling to the
bath \cite{escher_optimized_2011,*bensky_optimizing_2012,*petrosyan_reversible_2009,kofman_universal_2001,*kofman_unified_2004}.
The modulation rate must be faster than the inverse transfer time
$1/T$. On the other hand, decay or leakage of the single excitation
shared by the qubits and the bath must be either slower than $T$
or suppressed by an additional control field\textbf{ }\cite{clausen_bath-optimized_2010,*clausen_task-optimized_2012}. 

Among the diverse systems that are able to comply with these requirements,
we here suggest dipole-dipole (DD) coupled atoms \cite{petrosyan_scalable_2002,*opatrny_proposal_2003},
embedded in 1D photonic structures \cite{kurizki_resonant_1996,*kurizki_two-atom_1990,*friedler_long-range_2005,*shahmoon_strongly_2011,shahmoon_nonradiative_2013}.
Particulary appealing is a chain of atoms trapped just outside an
optical fiber whose dipole transition is within the optical bandgap
created by a grating in the fiber \cite{shahmoon_nonradiative_2013}.
If the dipole transition is just below the band edge, the DD coupling
is strongly enhanced while the radiative decay is suppressed by the
bandgap \cite{shahmoon_nonradiative_2013}. The resonance frequency
of the boundary atoms can be modulated faster than the DD couplings:
Modulation of the boundary-atom frequency shifts at a GHz rate, comparable
to the enhanced DD rate, should effectively control the transfer time
and fidelity along the chain in the presence of noise caused by sub-Kelvin
thermal fluctuations of the atomic positions and/or their random site
occupancy.

\textit{Conclusions}.---We have proposed a general, optimal and robust
dynamical-control of the tradeoff between transfer speed and fidelity
of qubit state transfer through a quantum channel in the presence
of either static or fluctuating noise. The only requirement for this
method to apply is for the channel to be symmetric with respect to
the source and target qubits and the number of eigenenergies has to
be odd. This leads to a central eigenvalue that is invariant against
static noise on the couplings, and have a gap separating it from adjacent
eigenenergies. Counterintuitively, we have shown that static noise
is more detrimental than fluctuating-noise, for a given noise strength
on the spin-spin couplings. Dynamical boundary-control has been used
to design an optimal spectral-filter that can minimize the leakage
to modes of the channel (here considered as a bath), that deteriorate
the transfer fidelity. The optimal filter is realizable by universal,
simple, modulation shapes that ensure the highest fidelity for a given
transfer time in both weak- and strong- coupling regimes, and are
robust against static and fluctuating noise on the spin-spin couplings.
As a result, the fidelity and/or the transfer time can be improved
by orders of magnitude compared with unmodulated transfer, while their
robustness against noise on the couplings is maintained or even improved.
The principles of this general treatment are extendable to other (non-Ising)
quantum channels as well.

We acknowledge the support of ISF-FIRST (Bikura) and the EC Marie
Curie (Intra-European) Fellowship (G.A.A.). 

\bibliographystyle{apsrev4-1}
\bibliography{Zwick-DCST}

\pagebreak{}

\begin{widetext} .

\section*{Supplementary material}

. **Supplemental information for "Dynamical control of state transfer through noisy quantum channels: optimal tradeoff of speed and fidelity"**

Analia Zwick, Gonzalo A. Alvarez, Guy Bensky and Gershon Kurizki

*Weizmann Institute of Science, Rehovot 76100, Israel*

## I. INTERACTION PICTURE REPRESENTATION OF THE HAMILTONIAN

The system-bath Hamiltonian (Eq. (3) of the main text) splits into a sum of symmetric and antisymmetric system operators that are coupled to bath odd- and even-modes: $H_{SB}(t) = \sum_{j=1}^{4} S_j \otimes B_j^{\dagger}$, where $S_{1(3)} = \alpha(t)(c_0 + (-)c_{N+1})$, $S_{2(4)} = S_{1(3)}^{\dagger}$, $B_{1(3)} = \sum_{k_{odd(even)}} \tilde{J}_k b_k$ and $B_{2(4)} = B_{1(3)}^{\dagger}$. In the interaction picture $H_{SB}(t)$ becomes

$$H_{SB}^I(t) = \sum_{j=1}^{4} S_j(t) \otimes B_j^{\dagger}(t), \tag{1}$$

where

$$S_{1(3)} = \alpha(t)(c_0 + (-)c_{N+1}), \; U_S(t) = e^{-i\int_0^t dt' H_S(t')}, \; B_j(t) = U_B^{\dagger}(t)B_j U_B(t), \; U_B(t) = e^{-iH_B t}; \tag{2}$$

and the evolution operators are

$$\begin{aligned}
U_S(t) = & \; |0\rangle_{SS}\langle 0| + \left(\frac{\cos(\sqrt{2}\phi(t))+1}{2}\right)(|0\rangle\langle 0| + |N+1\rangle\langle N+1|) + \left(\frac{\cos(\sqrt{2}\phi(t))-1}{2}\right)(|0\rangle\langle N+1| + |N+1\rangle\langle 0|) \\
& + \cos(\sqrt{2}\phi(t))|z\rangle\langle z| - i\frac{\sin(\sqrt{2}\phi(t))}{2}(|0\rangle\langle z| + |N+1\rangle\langle z| + h.c.) \\
U_B(t) = & \sum_{k=1, k\neq z}^{N} e^{-i\omega_k t}|k\rangle\langle k| + |0\rangle_{BB}\langle 0|,
\end{aligned} \tag{3}$$

where the states $|i\rangle = |0..01_i 0..0\rangle$ denote the one excitation subspace while $|0\rangle_S = |0_0 0_z 0_{N+1}\rangle_S$ and $|0\rangle_B = |0_1...0_N\rangle_B$ refer to the zero-excitation states in the system (S) and bath (B) respectively. Therefore, the bath operators are $B_{1(3)}(t) = \sum_{k_{odd(even)}} |\tilde{J}_k|^2 e^{-i\omega_k t}|k\rangle_B\langle 0|$, $B_{2(4)}(t) = B_{1(3)}^{\dagger}(t)$. We define a basis of operators $\hat{\nu}_i$ to describe the rotating system operators $S_j(t)$ via a rotation-matrix $\Omega_{j,i}(t)$, and they are given by

$$\begin{aligned}
\hat{\nu}_1 &= |0\rangle_S(\langle 0| + \langle N+1|) & \hat{\nu}_2 &= \hat{\nu}_1^{\dagger}, \\
\hat{\nu}_3 &= |0\rangle_S\langle z| & \hat{\nu}_4 &= \hat{\nu}_3^{\dagger}, \\
\hat{\nu}_5 &= |0\rangle_S(\langle 0| - \langle N+1|) & \hat{\nu}_6 &= \hat{\nu}_5^{\dagger},
\end{aligned} \tag{4}$$

such that $S_j(t) = \sum_i \Omega_{j,i}(t)\hat{\nu}_i$. Given that $S_1(t) = \dot{\phi}(t)\left(cos(\sqrt{2}\phi(t))\hat{\nu}_1 - i\sqrt{2}sin(\sqrt{2}\phi(t))\hat{\nu}_3\right)$, $S_3(t) = \dot{\phi}(t)\hat{\nu}_5$, $S_{2(4)}(t) = S_{1(3)}^{\dagger}(t)$, the rotation-matrix's vectors are

$$\begin{aligned}
\Omega_1(t) &= \dot{\phi}(t)\left(cos(\sqrt{2}\phi(t)), 0, -i\sqrt{2}sin(\sqrt{2}\phi(t)), 0, 0, 0\right) \\
\Omega_2(t) &= \dot{\phi}(t)\left(0, cos(\sqrt{2}\phi(t)), 0, i\sqrt{2}sin(\sqrt{2}\phi(t)), 0, 0\right) \\
\Omega_3(t) &= \dot{\phi}(t)(0, 0, 0, 0, 1, 0) \\
\Omega_4(t) &= \dot{\phi}(t)(0, 0, 0, 0, 0, 1).
\end{aligned} \tag{5}$$

## II. FIDELITY DERIVATION

From the system-bath interaction, Eq. (1), one can derive the system density matrix $\rho_S(t)$ in the interaction picture for a weak system-bath interaction as [1, 2]

$$\rho_S(t) = \rho_S(0) - t\sum_{i,i'} R_{i,i'}(t)[\hat{\nu}_i, \hat{\nu}_{i'}\rho_S(0)] + h.c., \tag{6}$$

where $R_{i,i'}(t) = \frac{1}{t} \sum_{j,j'} \int_0^t dt' \int_0^{t'} dt'' \Phi_{j,j'}(t'-t'') \Omega_{j,i}(t') \Omega_{j,i}^*(t'')$. The correlation between baths $j$ and $j'$ is denoted by $\Phi_{j,j'}(\tau) = \text{Tr}_B \left\{ B_j(\tau) B_{j'}(0) \rho_B(0) \right\}$ and $R_{i,i'}(t)$ is the average rate of change of the system's density matrix $\rho_S$ under the action of $[\hat{\nu}_i, \hat{\nu}_{i'} \rho_S(0)]$ caused by the bath. Using the basis $\nu_i$, we write the time-independent score matrix $\Gamma_{i,i'} = \langle \psi | [\hat{\nu}_i, \hat{\nu}_{i'} | \psi \rangle \langle \psi |] | \psi \rangle = \langle \hat{\nu}_i \hat{\nu}_{i'} \rangle - \langle \hat{\nu}_i \rangle \langle \hat{\nu}_{i'} \rangle$ [2], which describes the change of the fidelity with respect to the chosen basis $\hat{\nu}_i$.

Considering $|\psi\rangle = |100...0\rangle_{SB} = |1_0 0_z 0_{N+1}\rangle_S \otimes |0\rangle_B$ as initial state, the score matrix is reduced to

$$\Gamma_{i,i'} = \delta_{i,2} \delta_{1,i'} + \delta_{i,2} \delta_{5,i'} + \delta_{i,6} \delta_{1,i'} + \delta_{i,6} \delta_{5,i'}. \tag{7}$$

Then, the correlation functions in terms of the bath operators (described above) are

$$\Phi_{j,j'}(t-t') = \sum_{k_{odd}} |\tilde{J}_k|^2 e^{-i\omega_k(t-t')} \delta_{j,2} \delta_{1,j'} + \sum_{k_{even}} |\tilde{J}_k|^2 e^{-i\omega_k(t-t')} \delta_{j,4} \delta_{3,j'}, \tag{8}$$

and therefore

$$R_{i,i'}(T) = \frac{1}{T} \int_0^T dt \int_0^t dt' (\Phi_{2,1}(t-t') \Omega_{2,i}(t) \Omega_{1,i'}(t') + \Phi_{4,3}(t-t') \Omega_{4,i}(t) \Omega_{3,i'}(t')). \tag{9}$$

In the isolated 3-level system, perfect state transfer of the qubit-state $|1\rangle$ from the spin 0 (source qubit) to the $N+1$ (target qubit) occurs when the accumulated phase due to the modulation control $\phi(T) = \tilde{J}_z \int_0^T \alpha(t') dt'$ satisfies $\phi(T) = \frac{\pi}{\sqrt{2}}$. In the presence of the bath, the transfer fidelity of this qubit-state is given by $f_{0,N+1}(T) = |{_S}\langle \psi | \rho_S(T) | \psi \rangle_S|$ in the interaction picture and within the second-order approximation done in Eq. (6). It takes the form

$$f_{0,N+1}(T) = 1 - \zeta(T), \; \zeta(T) = T \times \Re\text{Tr} \left\{ \mathbf{R}(T) \mathbf{\Gamma} \right\}. \tag{10}$$

From Eqs. (7-9)

$$\zeta(T) = \int_0^T dt \int_0^t dt' \frac{\dot{\phi}(t) \dot{\phi}(t')}{\tilde{J}_z^2} (\Phi_{odd}(t-t') cos(\sqrt{2}\phi(t')) cos(\gamma\phi(t)) + \Phi_{even}(t-t')), \tag{11}$$

with $\Phi_{odd(even)}(\tau) = \Sigma_{k_{odd(even)}} \left| \tilde{J}_k \right|^2 e^{-i\omega\tau}$.

## III. EULER-LAGRANGE OPTIMIZATION

### A. Optimizing the modulation control $\alpha(t)$ for general non-Markovian gapped baths

In the energy domain, Eq. (11) has the form

$$\zeta(T) = \sum_{q=even,odd} \int G^q(\omega) F_T^q(\omega) d\omega, \tag{12}$$

where the Fourier transforms $G^q(\omega) = \mathcal{FT}(\Phi_q(\tau))$ and $F_T^q(\omega) = \mathcal{FT}(\frac{|\Omega_q(t)|^2}{2\pi})$ are the bath-spectrum and the filter-energy $q$ functions, respectively, for even or odd $q$. To determine the optimal modulation control, we minimize this overlap for a given $T$ by the variational Euler-Lagrange method. The shape of the bath-spectrum will change from channel to channel, but all of them have a common characteristic: a central gap around $\omega_z = 0$. Therefore, to find a general modulation control to minimize Eq. (12), we will assume a bath-spectrum that is continuous in the energy band with the exception of a central gap, and thus, we will maximize the filter function within this gap.

We maximize $F_T(\tau) = \int F_T(\omega) e^{-i\omega\tau} d\omega$, around $\omega_z = 0$, for every $\tau$ in order to assure the lowest frequency components of the filter-energy function under the accumulated phase $\phi(T) = J_z \int_0^T \alpha(t) dt$ and energy $E(T) = \tilde{J}_z^2 \int_0^T |\alpha(t)|^2 dt \geq \frac{\phi(T)^2}{T}$. The Euler-Lagrange equation is then

$$\frac{\partial F_T(\tau)}{\partial \alpha(t)} = \lambda_E \frac{\partial E(T)}{\partial \alpha(t)} + \lambda_\phi \frac{\partial \phi(T)}{\partial \alpha(t)}. \tag{13}$$

Since the desired sharp filter deals with the closest energies to 0, we focus on minimizing the overlap with $G^{even}(\omega)$. Given that $F_T^{s\,even}(\tau) = \int_0^T \alpha(t)\alpha(t+\tau)dt$, Eq. (13) becomes $\alpha(t+\tau) + \alpha(t-\tau) = \lambda_E \alpha(t) + \lambda_\phi$. For small $\tau$, it turns to be $\ddot{\alpha}(t) = -\tilde{\lambda}_E \alpha(t) + \tilde{\lambda}_\phi$ where $\tilde{\lambda}_E = \frac{-(\lambda_E - 2)}{\tau^2}$ and $\tilde{\lambda}_\phi = \frac{\lambda_\phi}{\tau^2}$ are the rescaled Lagrange multiplier. This differential equation has a general solution $\alpha(t) = A\sin(\omega_v t) + B\cos(\omega_v t) + C$, where the unknowns parameters will be optimized according to the required conditions, such us the boundary constraints, the transfer time, energy, etc. Relaxing the constraints and imposing only $\phi(T) = \frac{\pi}{\sqrt{2}}$, a condition on the frequency $\omega_v$ arises from the Fourier transform properties of the convolution between $\alpha(t)$ and the boxcar function on the time interval $[0, T]$.

The total filter will be low and flat outside a small range around 0 only if $\omega_v = \frac{\pi n}{T}$, $n \in \mathbb{Z}$, since the interference between the FT of the different terms of $\alpha(t)$ that oscillate with $\omega_v$ interfere destructively. On the other hand only if $n = 0, 1, 2$ the filter has a central and unique peak around 0 reducing the contribution of larger frequencies. For larger values of $n$, the central peak of the filter function is split and peaks at larger frequencies appear.

Therefore, the optimal solutions are found to be

$$\alpha_p(t) = \alpha_M \sin^p\left(\frac{\pi t}{T_p}\right),$$  (14)

with $p = 0, 1, 2$, $T_p = c_p \frac{\phi(T)}{\tilde{J}_z}$ and $c_p = \frac{\sqrt{\pi}\Gamma(\frac{1+p}{2})}{\Gamma(\frac{1+p}{2})}$ ($c_0 = 1$, $c_1 = \frac{\pi}{2}$, $c_2 = 2$).

### B. *Optimizing the modulation control $\alpha(t)$ for a specific non-Markovian bath*

The minimization of $\zeta(T)$ (11) can be also done for a specific bath-correlation function of a given channel. For example, for a finite homogeneous spin-channel, the exact correlation function of the bath is $\Phi_{odd(even)}(\tau) = \Sigma_{k_{odd(even)}} \left|\sqrt{\frac{2}{N+1}}J\sin(\frac{k\pi}{N+1})\right|^2 e^{-i2J\cos(\frac{k\pi}{N+1})\tau}$ and has recurrences and time fluctuations due to mesoscopic revivals, while at short times $t$, it behaves as a Bessel function $\Phi(t) = \frac{2(\alpha_0 J)^2}{j\tau}\mathcal{J}_1(2Jt)$. The latter correlation function represents the limiting case of an infinite channel and it gives a continuous bath-spectrum that becomes a semicircle. In the case of a finite channel, $G(\omega)$ will be discrete but modulated by the semicircle with a central gap. If disorder is considered, the position of the spectrum lines fluctuates from channel to channel but they are essentially modulated by the semicircle with a central gap as was considered in the Fig. 1 of the main text, where $G(\omega) = \frac{1}{2}\sqrt{4J^2 - \omega^2}(1 - \Theta(\omega - \omega_l)\Theta(\omega + \omega_l))$, $\omega_l = \frac{3\omega_{z+1}}{4}$. This is the Wigner-distribution for fully randomized channels [3, 4] with a central gap.

Once the specific channel and $\Phi_{odd(even)}(\tau)$ are given, the Euler-Lagrange method can be implemented as follow. The minimization of $\zeta(T)$ (Eq. 11) can be done under a constraint $\chi(T)$ to avoid unphysical results [1, 2, 5, 6]. The Euler-Lagrange equation turns then

$$\frac{d}{dt}(\frac{\partial \zeta}{\partial \dot{\phi}} - \lambda \frac{\partial \chi}{\partial \dot{\phi}}) - (\frac{\partial \zeta}{\partial \phi} - \lambda \frac{\partial \chi}{\partial \phi}) = 0,$$  (15)

where $\lambda$ is the Lagrange multiplier factor.

Choosing the energy as a constraint $\chi(T) = E(T) = \tilde{J}_z^2 \int_0^T \dot{\phi}^2(t)dt$, the optimal modulation is given by the integro-differential equation

$$\ddot{\phi}(t) = \frac{\sqrt{E}Q(t, \phi(t), \dot{\phi}(t))}{\tilde{J}_z\sqrt{\int_0^T dt\left|\int_0^t dt' Q(t', \phi(t'), \dot{\phi}(t'))\right|^2}},$$  (16)

where

$$Q(t, \phi(t), \dot{\phi}(t)) = \frac{1}{2\tilde{J}_z^4}(\int_0^T dt' \Theta(t - t')\dot{\phi}(t')\left(\frac{d\Phi_{odd}(t-t')}{dt}\cos(\sqrt{2}\phi(t))\cos(\sqrt{2}\phi(t')) + \frac{d\Phi_{even}(t-t')}{dt}\right) + \dot{\phi}(t)\left(\Phi_{odd}(0)\cos^2(\sqrt{2}\phi(t)) + \Phi_{even}(0)\right)).$$  (17)

Eq. (16) should satisfy the boundary conditions $\phi(0) = 0$ and $\phi(t = T) = \frac{\pi}{\sqrt{2}}$ to ensure the required state transfer.

### C. *Optimizing the modulation control $\alpha(t)$ for a Markovian Bath*

For a Markovian bath, the infidelity function (11) to be minimized becomes

$$\zeta(T) = \Re \int_0^T dt \frac{\dot{\phi}^2(t)}{\tilde{J}_Z^2}\left(\Phi_{odd}(0)\cos^2(\sqrt{2}\phi(t)) + \Phi_{even}(0)\right).$$  (18)

Under the Euler-Lagrange method with the energy constraint, the differential equation is obtained

$$\ddot{\phi}(t)\left(\Phi_{odd}(0)cos^2(\sqrt{2}\phi(t)) + \Phi_{even}(0) - 2\lambda\tilde{J}_z^2\right)$$
$$-\gamma\dot{\phi}^2(t)\,\Phi_{odd}(0)cos(\sqrt{2}\phi(t))sin(\sqrt{2}\phi(t)) = 0. \tag{19}$$

This equation has a non-trivial analytical solution and the modulation which minimize $\zeta(T)$ is given by the transcendental equation

$$T\int_0^{\phi(t)}\sqrt{cos(2\gamma\varphi)\Phi_{odd}(0)_{k_{odd}} + \Phi_{odd}(0) + 2\Phi_{even}(0) - 2\lambda\tilde{J}_z^2}d\varphi$$
$$-t\int_0^{\phi(T)}(\sqrt{2(\Phi_{odd}(0)cos^2(\gamma\varphi) + \Phi_{even}(0) - \lambda\tilde{J}_z^2)}d\varphi = 0.$$

Without constraint ($\lambda = 0$), the optimal modulation can be approximated by $\alpha(t) \approx \alpha_M(a + b\,sin^p(\frac{t\pi}{T}))$, where $p \sim 3.5$, $\frac{a}{b} \sim \frac{1}{3}$ and $\alpha_M = \max\alpha(t)$. The infidelity for this optimal modulation almost coincides with the one obtained without modulation $1 - F(T) \approx \frac{\pi^2 N}{6\sqrt{2}JT}(1 - \frac{\pi^2 N}{16\sqrt{2}JT})$ with $T \approx \frac{\pi\sqrt{N}}{2\alpha_M J}$, and they only differ by about 0.1%.

## IV. IMPLEMENTATION

The boundary coupling strengths can be engineered as a function of time, using Trotter-Susuki decompositions, by a sequence of $\pi$-pulses applied only at the boundary spins at suitable times; for example, by a series of cycles of duration $\tau_c \ll J_i$ that contain two $\pi$-pulses separated as $\tau_1^j - \pi - \tau_2^j - \pi - \tau_3^j$, where $\tau_1^j + \tau_2^j + \tau_3^j = \tau_c$ and $j$ represents the cycle number. In this way, the modulation control $\alpha(t)$ (in Eq. (1) from main text) for $t = j\tau_c$ will be given by $\alpha(j\tau_c) = (\tau_1^j - \tau_2^j + \tau_3^j)/\tau_c$. One can modulate $0 \le \alpha(j\tau_c) = (\tau_1^j - \tau_2^j + \tau_3^j)/\tau_c \le 1$ as a function of the time $j\tau_c$ as needed, by choosing appropriate values of $\tau_i^j$ at every cycle $j$.

\end{widetext}
\end{document}